# Visualizing reacting single atoms in chemical reactions, advancing the frontiers of materials research

## Edward D. Boyes and Pratibha L. Gai

Heterogeneous gas–solid catalyst reactions occur at the atomic level, and understanding and controlling complex catalytic reactions at this level is crucial for the development of improved processes and materials. There are postulations that single atoms and very small clusters can act as primary active sites in chemical reactions. Early applications of our novel aberration-corrected (AC) environmental (scanning) transmission electron microscope (E(S)TEM) with single-atom resolution are described. This instrument combines, for the first time, controlled operating temperatures and a continuous gas environment around the sample with full AC STEM capabilities for real-time *in situ* analysis and visualization of single atoms and clusters in nanoparticle catalysis. ESTEM imaging and analysis in controlled gas and temperature environments can provide unique insights into catalytic reaction pathways that may involve metastable intermediate states. Benefits include new knowledge and more environmentally friendly technological processes for health care and renewable energy as well as improved or replacement mainstream technologies in the chemical and energy industries.

## Introduction

Reliable *in situ* observations of single atoms in gas–solid catalyst reactions under controlled reaction conditions of gas environments and elevated temperatures is a key goal in understanding heterogeneous gas–solid catalytic reactions.[1] Recent applications exploit the development of atomic-resolution environmental (scanning) transmission electron microscopy (E(S)TEM) for understanding the role of gas–surface interactions in nanoscale catalysts in their functioning state. Temperature, time, and pressure-resolved studies on real systems have been realized, with the aim of bridging pressure and materials gaps at operating temperatures.[2,3]

The use of surface science techniques in an ultra-high vacuum, with extended single crystal surface model systems to understand fundamental aspects of heterogeneous catalysis, has been invaluable. However, direct industrial applications of these studies have been limited by the pressure and materials gaps that exist between them and industrial applications. In addition, practical heterogeneous catalyst surfaces are generally not extended as perfect single crystal faces but the much more complex shapes of nanoparticles on supports.[4,5] Early environmental TEM (ETEM) methods also did not provide atomic resolution.[6]

## Atomic-resolution ETEM

In developing the first atomic-resolution ETEM,[2,3] we took a new approach to organize and design an instrument dedicated to environmental cell (ECELL or gas reactor system) operations with continuously flowing gas and elevated specimen temperatures. The ECELL facilities are integral to the electron microscope, which has been important in the development of the atomic-resolution ETEM for probing *in situ* gas–solid reactions directly at the atomic level in real time, under controlled gas atmosphere and temperature conditions, and for *in situ* nanosynthesis.[3] The whole electron microscope column in a modern high-resolution (S)TEM has been modified for the ECELL functionality, rather than only the immediate region around the sample.

Highlights of this development have included objective lens pole pieces incorporating radial holes for the critical first

Edward D. Boyes, University of York; ed.boyes@york.ac.uk
Pratibha L. Gai, University of York; pratibha.gai@york.ac.uk

stage of differential pumping.[3] The basic geometry is a four-aperture system with apertures innovatively mounted inside the bores of the objective lens pole pieces. The regular electron microscope sample chamber is used as the controlled reaction ECELL or reactor and thus, is integral to the core instrument.[3] The environmental transmission electron microscope can be operated either with gas environments or as a conventional high-vacuum transmission electron microscope without compromising the atomic-resolution imaging. This system permits high gas pressures in the ECELL sample region while maintaining high vacuum in the rest of the environmental transmission electron microscope. A gas manifold system constructed from stainless steel modules controls the inlet of flowing gases into the ECELL of the microscope. The sample is in a controlled continuous flowing gas environment similar to conditions in technological reactors, with a positive flow rate of gas. A hot stage allows samples to be heated routinely at controlled temperatures up to about 1000°C (or higher with less control using special holders), and a mass spectrometer is included for inlet gas analysis.

For dynamic atomic resolution studies, up to a few millibars (mbar) or several hundred Pa of gas pressure are typically used in the ECELL, but our original version can operate with gas-type dependent pressures >50 mbar (5 kPa). Higher gas pressures compromise the resolution due to multiple scattering effects of the electron beam through denser gas layers, and the extent of this depends on the type of gas. Electronic image shift and drift compensation help stabilize high-resolution images for data recording (with a time resolution of the order of 1/30 s or better), on film, on charge-coupled devices, or with the originally used low-light television video system incorporating real-time recursive digital processing.

Minimally invasive low-dose electron beam techniques are used throughout. *In situ* data under low electron dose conditions are checked in parallel blank calibration experiments, in which dynamic experiments are conducted without the electron beam, and the beam is switched on only to record the final reaction endpoint. The aim is non-invasive characterization under benign conditions without contamination.[1–3,7,8] By providing the gas supply and some other facilities as part of the microscope, this arrangement allows use of a wide range of existing designs of specimen holders for flexibility, simplicity, economy, and reduced set-up time and cost in new scientific applications. It is essential for the specimens to be clean, both to ensure that the intended chemistry is being probed and to avoid contamination build up in the electron microscope. A hot stage (rather than uncontrolled and highly profiled beam heating) must be used to heat the sample. Because of the small amounts of solid reactant in the EM sample, measurement of reaction products are carried out on larger samples in an *ex situ* microreactor operating under similar reaction environments for the nanostructural correlation with reactivity.

In catalysis, the correlation of the atomic structure and function is crucial in optimizing synthesis and developing improved materials and processes. Under carefully controlled conditions, data from *in situ* ETEM can be directly related to structure–activity/selectivity/stability relationships in technological processes. The smallest particles (clusters) or even single atoms may be the most active and selective catalysts, but they may not be very stable, leading to an early decrease in catalyst performance and therefore process profitability.

The development of the atomic-resolution ETEM[3] has opened up a new field for studying gas–solid reactions at working temperatures at the atomic level.[4,5] Commercial versions of this development are now widely used by numerous researchers globally.[7–13] The latest versions can support controlled reaction environments of temperature and gas involving pressure and time-resolved studies with <0.1nm resolution and picometer precision.[14]

### Benefts of aberration correction for *in situ* E(S)TEM studies

For enhancing the atomic-resolution ETEM, we developed the first double aberration corrected (AC) E(S)TEM in order to visualize single atoms in catalytic reactions under controlled continuous flowing gas atmospheres at working temperatures. It extends high vacuum and ETEM analyses with full analytical electron microscopy facilities. These include controlled conditions of a continuous gas environment and initial operating temperatures >650°C with unrestricted high-angle annular dark field (HAADF) Z-contrast ESTEM imaging of single atoms and clusters, low background energy-dispersive x-ray (EDX) spectroscopy, AC ESTEM electron energy loss spectroscopy, and wide-angle electron diffraction analyses of nanoparticle structures and crystallography in dynamic *in situ* experiments. HAADF is also required to set up the STEM corrector[15] introduced for the AC ESTEM system. A schematic diagram (**Figure 1**) outlines the basic configuration of the ESTEM.

Aberration correction is particularly beneficial in dynamic *in situ* experiments because it enables the extraction of the maximum possible information from each single image or other data in a dynamic sequence.[1] Based on these considerations, we proposed the AC ESTEM.[16] Calibration experiments to ensure minimally invasive conditions are performed according to procedures described in the previous sections and to avoid secondary effects such as contamination.

The need to accommodate special specimen holders in an AC instrument has been one criterion driving our specification of the JEOL 2200FS field emission gun (FEG) TEM/STEM double aberration-corrected (2AC) electron microscope operating at 200 kV in the Nanocentre at the University of York. Currently, all the AC ETEMs use the earlier atomic-resolution ETEM design of Gai and Boyes[2,3] described previously.[7]

Since aligning the sample into a zone axis is a prerequisite for atomic-resolution electron microscopy of crystals, in addition to a hot stage, an increased specimen tilt range is also desirable. Both of these conditions benefit from the larger gap (HRP/Midi) objective lens pole piece[1,17] shown in **Table I**.

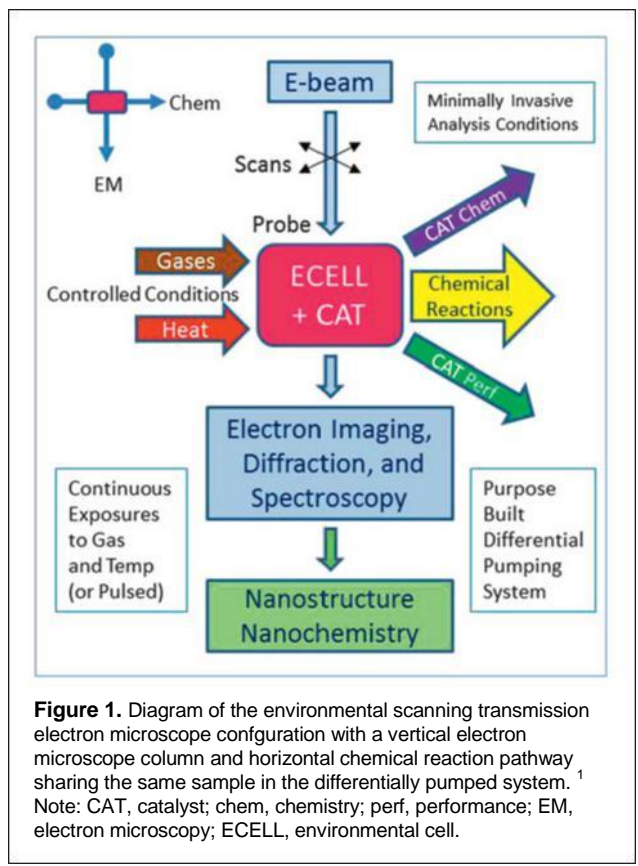

**Figure 1.** Diagram of the environmental scanning transmission electron microscope configuration with a vertical electron microscope column and horizontal chemical reaction pathway sharing the same sample in the differentially pumped system.[1] Note: CAT, catalyst; chem, chemistry; perf, performance; EM, electron microscopy; ECELL, environmental cell.

The Cs (C3) CEOS (Corrected Electron Optical Systems GmbH) aberration correctors [15] for both STEM probe forming and TEM imaging in the instrument were used to provide the desired specimen geometry with minimal effect on the <0.1 nm resolution of the system. The effect of the increased pole piece gap in the range of 2–5 mm on the spherical aberration coefficient ($C_s$) is more evident (x2) than on the chromatic aberration coefficient $C_c$ (x1.2) and of course $C_s$ is now corrected ($C_s$ ― 0). This means the performance is largely retained with the increased space and ensuing operational flexibility.

The advantages of the configuration we have adopted include: (1) promoting an HRTEM contrast transfer function extending to higher spatial frequencies and resolution in the data; (2) allowing image recording at close to zero defocus to strengthen interpretation of information from small nanoparticles and clusters on supports; (3) using high-resolution TEM as well as high-resolution STEM; (4) facilitating HAADF STEM; (5) and extending HAADF STEM resolution to ≤ 0.1 nm.[1] The 2AC ESTEM pumping system at York is configured to be gas tolerant by using a combination of turbomolecular and molecular drag pumps for the main column evacuation systems.[1,17] Ion pumps continue to be used (and added) pre-FEG and in the field emission electron gun itself.[17] Despite the extensive *in situ* modifications to the system, single-atom imaging has been possible in practical catalysts under reaction conditions.[18,19]

The first applications of the 2AC ESTEM, modified in York to provide both ETEM,[1] and for the first time ESTEM, with full analytical functionalities have been reported earlier.[18,19] The primary applications to date have been the imaging of single Pt atoms showing their existence on a carbon supported model Pt nanoparticle catalyst,[18,19] and the combined use of the imaging and analytical systems for characterizing core–shell diesel exhaust catalysts. TEM and STEM aberration correction montage procedures (Zemlin tableaux [15] and image resolution to better than 0.1 nm) have been maintained with the hot stage and gas environments in the *in situ* 2AC ESTEM and the samples held at elevated temperatures >500°C.[18,19] Details of this approach have been published previously.[1,",19] TEM $C_s$ correction also allows use of the previously unattainable combination of conditions with small negative $C_s$ and positive defocus [20] to create a new high-contrast imaging mode for atomic-resolution studies. For heating, it has been possible to modify the original Gatan 628 furnace hot stage design,[18] retaining the 3 mm disc specimen capability, with new electronics for 0.1 nm STEM operations. Microelectromechanical systems (MEMS) hot stage technology from DENS solutions, with on-chip temperature measurement, also provides high lateral stability with much shorter response times and is attractive for particulate catalyst samples placed directly onto the filmed windows.

In contrast to the usual high-vacuum STEM analytical environment, the AC E(S)TEM system operates in both ESTEM and ETEM modes with 1 to >10 Pa gas pressures for a gas supply of up to 100,000 monolayers per second (>0.1 mbar) measured directly at the sample position with inlet gas pressures up to a few mbar or 100s of Pa. Even at more modest Pa pressures, the gas supply should generally still be adequate to flood the sample surface with gas coverage. The system supports analysis of key intermediate catalyst phases that may be metastable with respect to the gas atmosphere, temperature, and time. They may be critical to understanding catalyst mechanisms and to designing new ones,[1,17] but they are not reliably accessible with *ex situ* or high-vacuum methods of analysis.

**Table I.** Summary of some high-performance objective lens pole piece parameters for high-resolution electron microscopy.

| Polepiece Type | Gap Range (mm) | Uncorrected $C_s$ (mm) | Uncorrected $C_c$ (mm) | $C_c$ with $C_s$ AC | Std Hot Stage Fits? |
|---|---|---|---|---|---|
| UHR | 2.2–2.5 | 0.5–0.6 | 1.0–1.2 | 1.4 | No |
| Midi | 4.3–5.4 | 1.0–1.2 | 1.2–1.4 | 1.6 | Yes |

Note: UHR, ultra high resolution; Midi describes pole pieces accommodating higher specimen tilt while still providing good resolution (see text); $C_s$ and $C_c$ are, respectively, the spherical and chromatic aberration coefficients, which without correctors, control microscope performance. Adapted with permission from Reference 1.

## AC ESTEM of Pt and Au nanocatalysts

We describe initial experiments using ESTEM of supported Pt nano-catalysts of a type widely used in catalyst technologies and fuel cells as distributed video signals and <0.1 nm with expanded contrast and shifted black level. In AC ETEM images of similar areas, it is difficult to reliably discern the single-atom positions.[19] With the ESTEM, the complete population of single atoms has been identified and migratory single atoms on the support between the more substantial clusters have been imaged for the first time by the authors under controlled gas and temperature environments.[18] This observation of reacting single atoms is important in understanding the mechanisms of catalytic reactions and deactivation processes.

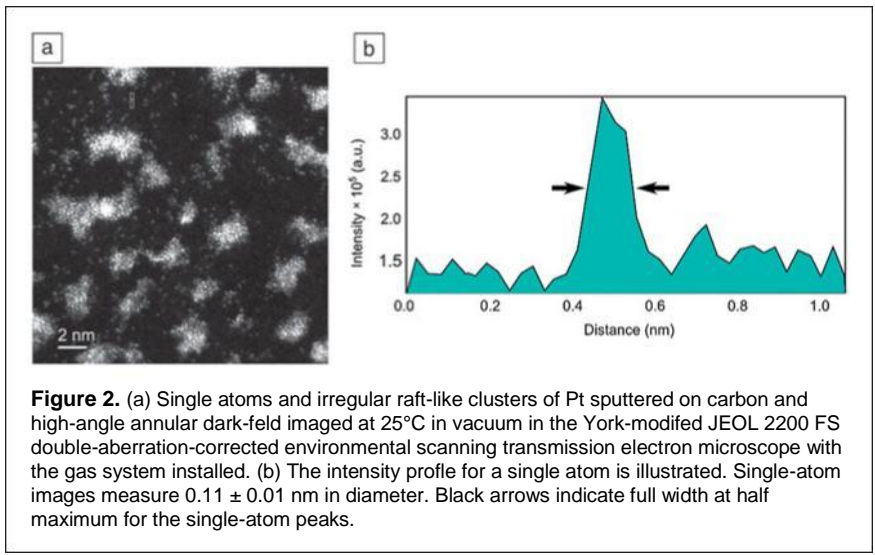

**Figure 2.** (a) Single atoms and irregular raft-like clusters of Pt sputtered on carbon and high-angle annular dark-feld imaged at 25°C in vacuum in the York-modifed JEOL 2200 FS double-aberration-corrected environmental scanning transmission electron microscope with the gas system installed. (b) The intensity profile for a single atom is illustrated. Single-atom images measure 0.11 ± 0.01 nm in diameter. Black arrows indicate full width at half maximum for the single-atom peaks.

**Figure 3**a[18] shows a sequence of single Pt atoms and clusters, some of which are seen to be substantially modified and evolving into well-ordered three-dimensional (3D) single crystal particles (such as A in Figure 3b ), on a carbon support in an atmosphere of hydrogen gas at 25°C. Under these conditions, even at 25°C, the internal structure of some of the particles becomes highly ordered with a strong (110) texture and external surfaces faceting into {111} faces. In Figure 3b , at 400°C in hydrogen, in addition to more ordered particles, there is clear evidence of residual single atoms still present on the support away from the nanoparticles. Figure 3c shows further development of nanoparticle structures imaged in hydrogen at 500°C after 30 minutes of exposure to the reaction conditions of gas and temperature. This image shows far fewer distinct single atoms recorded, and their typical contrast is greatly reduced. There may be questions of mobility and image sampling frequency rather than thermal diffuse contrast, or they may report an actual absence of loose single atoms. More 3D particles are observed at higher temperature compared to the initial observations of as-deposited

energy sources. Applications include direct *in situ* studies of single-atom migration in a model supported metal particle heterogeneous catalyst under controlled gas and temperature operating conditions related to industrial applications. The samples for the initial studies reported here were prepared by sputtering Pt onto a carbon film support. They were mildly plasma cleaned, and sample contamination issues were addressed by using a hot stage. The full sample handling procedures are described elsewhere.[18]

The images in this section are all HAADF images obtained using the original JEOL (upper) STEM detector. The recording conditions for the STEM data included magnifications of 8–12 Mx with 512 and 1024 line frames and sampling at 0.015 or 0.03 nm; that is, at 6x or 3x over-sampling of the 0.1 nm digital imaging resolution element, using pixel dwell times of 19 or 38 μs with 50 Hz line synchronization in Europe (equivalent line times in North America or elsewhere with 60 Hz mains power would be different). Some images were also recorded with faster frame integrations. The incident probe had a calibrated 24 mrad convergence semi-angle, and the collection angles for HAADF imaging were 110 mrad and 170 mrad, respectively, for the inner and outer diameters of the detector. Short sequences of images recorded at reduced frame times and pixel counts were also assembled into a video to analyze the dynamic, and in some cases competitive, nature of the processes involved.

**Figure 2**[18] shows single atoms and clusters of Pt on a carbon support with a corresponding HAADF AC ESTEM image intensity profile of a single atom recorded at 25°C. The very thin (1–2 atoms high), raft-like clusters are made up primarily of partially ordered {111} spacings with a general (110) texture. The full width at half maximum for the single-atom image peaks are 0.11 ± 0.01 nm in the full range

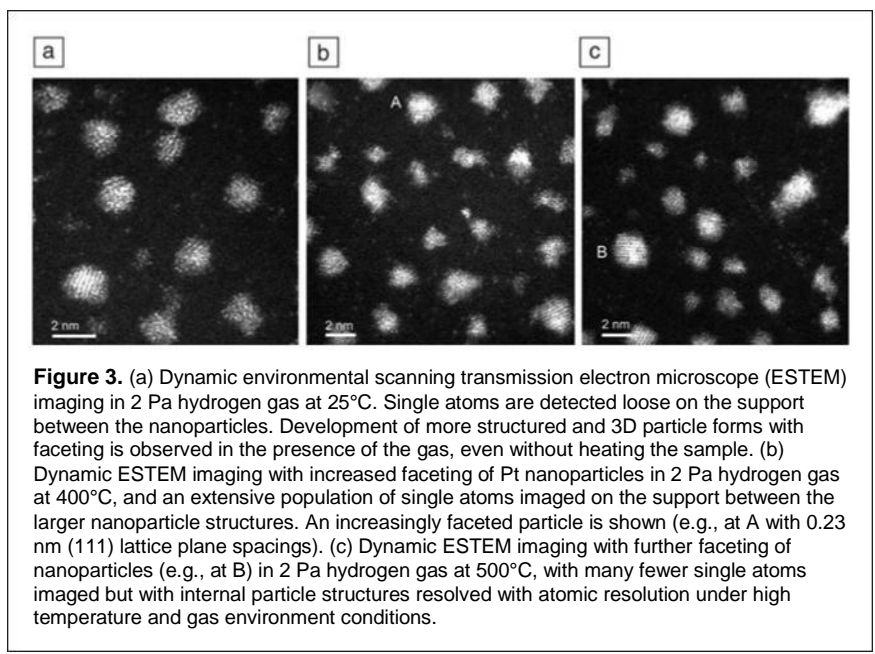

**Figure 3.** (a) Dynamic environmental scanning transmission electron microscope (ESTEM) imaging in 2 Pa hydrogen gas at 25°C. Single atoms are detected loose on the support between the nanoparticles. Development of more structured and 3D particle forms with faceting is observed in the presence of the gas, even without heating the sample. (b) Dynamic ESTEM imaging with increased faceting of Pt nanoparticles in 2 Pa hydrogen gas at 400°C, and an extensive population of single atoms imaged on the support between the larger nanoparticle structures. An increasingly faceted particle is shown (e.g., at A with 0.23 nm (111) lattice plane spacings). (c) Dynamic ESTEM imaging with further faceting of nanoparticles (e.g., at B) in 2 Pa hydrogen gas at 500°C, with many fewer single atoms imaged but with internal particle structures resolved with atomic resolution under high temperature and gas environment conditions.

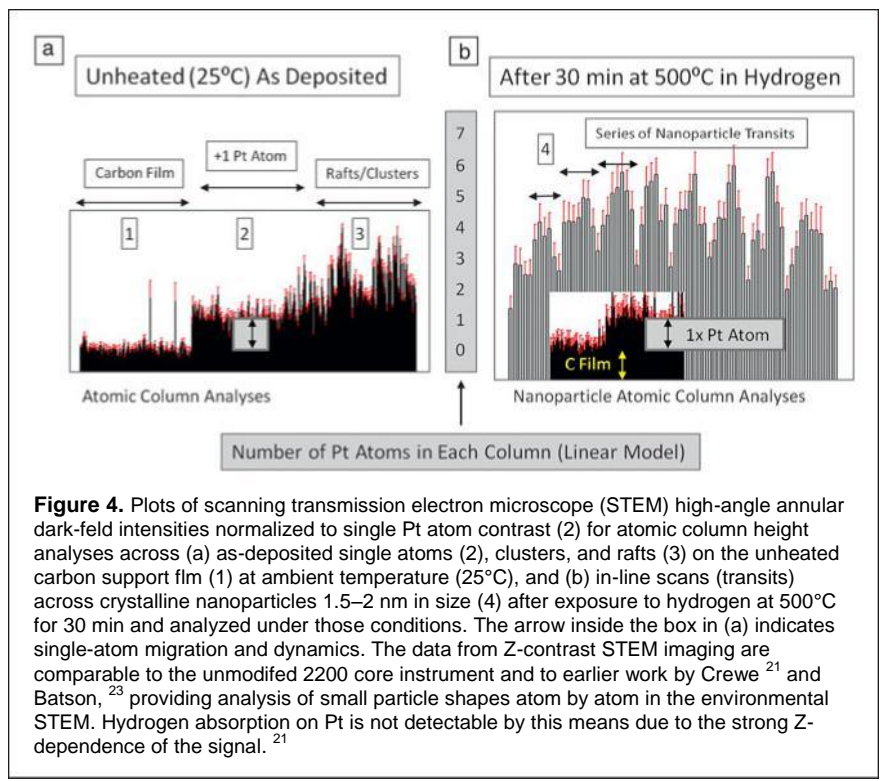

**Figure 4.** Plots of scanning transmission electron microscope (STEM) high-angle annular dark-field intensities normalized to single Pt atom contrast (2) for atomic column height analyses across (a) as-deposited single atoms (2), clusters, and rafts (3) on the unheated carbon support film (1) at ambient temperature (25°C), and (b) in-line scans (transits) across crystalline nanoparticles 1.5–2 nm in size (4) after exposure to hydrogen at 500°C for 30 min and analyzed under those conditions. The arrow inside the box in (a) indicates single-atom migration and dynamics. The data from Z-contrast STEM imaging are comparable to the unmodifed 2200 core instrument and to earlier work by Crewe [21] and Batson, [23] providing analysis of small particle shapes atom by atom in the environmental STEM. Hydrogen absorption on Pt is not detectable by this means due to the strong Z-dependence of the signal. [21]

raft-like structures before exposure to hydrogen. They are accompanied by more faceted particle perimeters based on regular single crystal internal forms for the Pt particles. The images illustrate the tension that exists between faceted nanoparticles, which have inherent stability, and irregular ones with a higher proportion of low coordination more reactive surface atoms.

Crewe et al. [21] introduced quantitative measurements of a restricted number of atoms ($n$) in each column (up to $n$—10 atoms per column) using Z-dependent analytical HAADF electron imaging. With superior instruments, the sensitivity of these analyses has improved. [22] Here, we identify single atoms in a similar way as Batson et al. [23] with preliminary illustrative data in **Figure 4**.[18] They show discrete levels in the video signals from single atoms, with multiple ($n$ up to 4 but typically 1–2) layers of atoms in the initial clusters, based on a linear model [21] for intensity with $n$ (for $n$ <10). The particles more developed at higher temperatures have been analyzed by the video intensity of discrete columns of atoms in a regular crystal structure (Figure 3), and they show a thicker form of particle ($n$ up to 8 ± 1 measured) than the flatter, more raft-like, and disordered forms of the initial clusters ($n$ = 1–2) (Figure 3a).

*In situ* hydrogen reduction in Figure 3b indicates the presence of fewer single atoms on the support than in Figure 3a, with clusters and increased faceting of the nanoparticles. This indicates that the nanoparticles act as a source of potential adatoms and clusters and as recipients for them. In hydrogen reducing environments, low coordination surface atoms are replaced by surface facets through local rearrangements to minimize surface energy, [18,19] further illustrated in **Figure 5**. Whilst overall loss of surface area and activity is often the result of Ostwald ripening, the existence of single atoms and small clusters on the surface of the support during the intermediate stages has important implications for catalysis and the role of nanoparticles generally. Through the use of a support material with an abundance of anchoring sites, migrating atoms may be stabilized and provide further active sites for adsorption, with the nanoparticles acting as reservoirs and recipients of adatoms and migratory clusters.

In addition to maintaining the catalyst surface area by preventing deactivation from sintering processes, understanding the evolution of the identified active sites at single-atom resolution is also needed. The new knowledge of single-atom behavior should lead to the development of more active and more economically and environmentally attractive catalysts.

## Applications of AC ETEM to biofuels

Alternatives to fossil fuels are being sought in order to reduce the world's dependence on non-renewable energy resources. Biofuels can be derived from plant oil feedstocks and biological materials. Many countries have set targets for partial replacement of fossil fuels with renewable biofuels. Biodiesel is one example of a clean and renewable biofuel. Solid catalysts (e.g., alkaline earth oxides) offer advantages, especially pertaining to biodiesel separation and the opportunity for continuous process operation. Although solid catalysts have great promise in plant oil triglyceride transesterification to biodiesel, the identification of active sites and surface nanostructures during processing is required for the development of energy

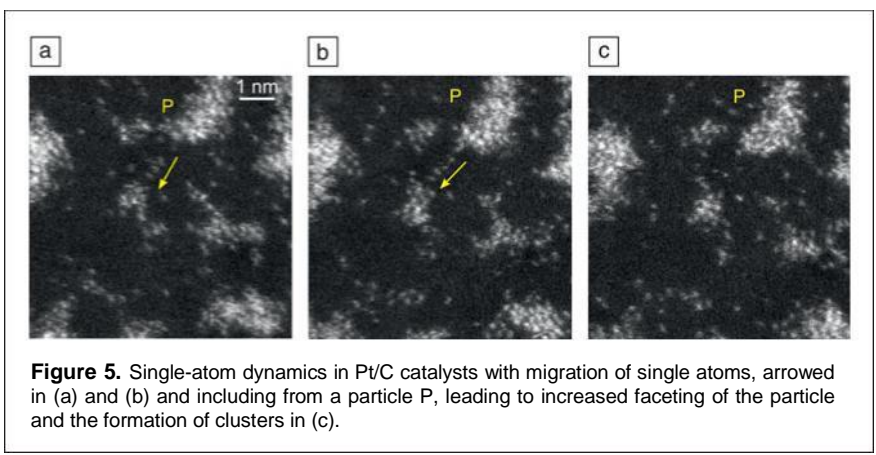

**Figure 5.** Single-atom dynamics in Pt/C catalysts with migration of single atoms, arrowed in (a) and (b) and including from a particle P, leading to increased faceting of the particle and the formation of clusters in (c).

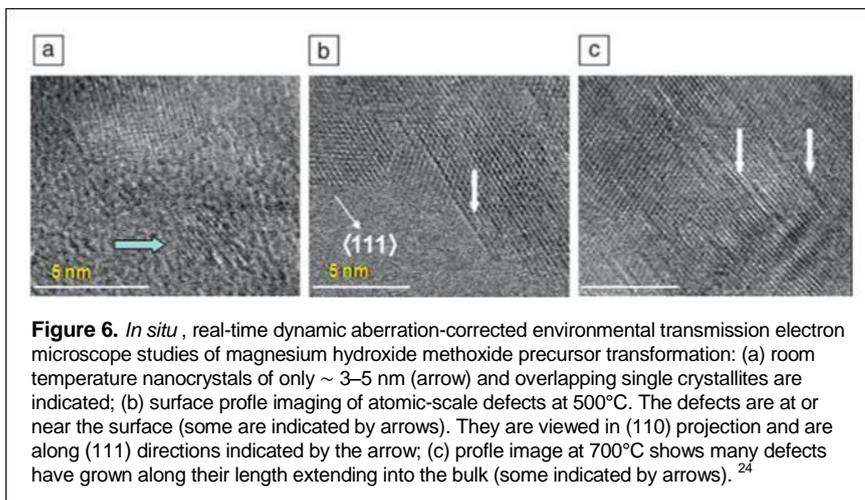

**Figure 6.** *In situ*, real-time dynamic aberration-corrected environmental transmission electron microscope studies of magnesium hydroxide methoxide precursor transformation: (a) room temperature nanocrystals of only ~ 3–5 nm (arrow) and overlapping single crystallites are indicated; (b) surface profile imaging of atomic-scale defects at 500°C. The defects are at or near the surface (some are indicated by arrows). They are viewed in (110) projection and are along (111) directions indicated by the arrow; (c) profile image at 700°C shows many defects have grown along their length extending into the bulk (some indicated by arrows).[24]

## Acknowledgments
This work is supported by the EPSRC (UK) critical mass grant EP/J0118058/1 awarded to PLG and EDB. The authors thank Michael Ward, Leonardo Lari, and Ian Wright for support.

efficient catalysts for biofuel synthesis.[24] Alkaline earth oxides, such as MgO, are of technological interest in plant triglyceride transesterification. It adopts a cubic structure with $a = 0.4212$ nm, where $a$ is the lattice parameter.

High-quality nanoscale MgO practical powder catalysts can be prepared using magnesium hydroxide methoxide precursor $(Mg(OH)(OCH_3))$.[24] The dynamic atomic level transformation of the precursor can be followed in real time under controlled calcination conditions using *in situ* atomic-resolution AC ETEM. The observations of the catalyst nanostructure are quantified with parallel studies of the catalyst performance and physico-chemical studies.[24]

The dynamic precursor transformation is illustrated in **Figure 6**.[24] Figure 6a shows randomly orientated, overlapping single crystallites of only 3–4 nm in an amorphous medium. The crystallites are primarily in (110) and (001) orientations at 25°C. In Figure 6b, atomic-scale defects are observed at the catalyst surface at 500°C. Finally, growth of the defects is observed at 700°C in Figure 6c. Analysis of defects indicated that they are partial screw dislocations formed by a glide shear mechanism along the (111) direction, and the shear or the displacement vector **R** of the type $\pm a/2$ (111) lies in the shear plane. The defects are associated with coplanar anion vacancies in the active sites of the catalyst in the transesterification of plant triglycerides to biodiesel. The direct linear correlation between the surface atomic-scale glide defect density and improved catalytic performance has offered a new route to energy efficient catalysts for biofuel synthesis.

## Conclusion
The visualization of single atoms reacting in controlled, reduced environments at working temperatures revealed by the new AC ESTEM suggests that low coordination surface atoms and loose single atoms are key to understanding the performance in gas–solid catalyst reactions. AC ETEM studies of catalysts for biofuels suggest that atomic-scale glide defects are important in the development of energy efficient catalysts.